\documentclass[         
aps,                    
prd,                    
showpacs,               
nofootinbib,            
showkeys,               %
preprintnumbers,        %
floatfix]               
{revtex4}               
\usepackage{graphicx,longtable}
\usepackage{color}
\begin{document}

\title{Collective neutrino oscillations in matter and CP violation}

\author{J\'er\^ome Gava}
\affiliation{Institut de Physique Nucl\'eaire, F-91406 Orsay cedex,
France}

\author{Cristina Volpe}
\affiliation{Institut de Physique Nucl\'eaire, F-91406 Orsay cedex,
France}

\date{27th June 2008}

\begin{abstract}
We explore CP violation effects on the neutrino propagation in dense environments, such as in core-collapse supernovae, where the neutrino self-interaction induces non-linear evolution equations. We demonstrate that the  electron (anti)neutrino fluxes
are not sensitive to the CP violating phase if the muon and tau neutrinos interact similarly with matter. 
On the other hand,
we numerically show that new features arise, because of the non-linearity and the flux dependence of the evolution equations,
when the muon and tau neutrinos have different fluxes at the neutrinosphere
(due to loop corrections or of physics beyond the 
Standard Model).
In particular, the electron (anti)neutrino probabilities and fluxes depend upon the CP violating phase.
We also discuss the CP effects induced by radiative corrections to the neutrino refractive index.  
\end{abstract}



 \maketitle


\section{Introduction}
One of the major open issues in neutrino physics is the possible existence 
of CP violation. As with the recent crucial experimental discoveries on neutrino
oscillations, the answer to this question has fundamental implications
in high-energy physics, astrophysics and cosmology e.g. to understand the 
matter versus anti-matter asymmetry in the Universe. The observation
that the weak interaction violates the CP symmetry in the quark sector was first established in 1964
\cite{Christenson:1964fg}. Future strategies to search for CP violation in the lepton sector depend upon 
the actual value of one yet unknown neutrino oscillation parameter, i.e. $\theta_{13}$ \cite{Ardellier:2006mn}, and require 
long term accelerator projects producing very intense neutrino beams \cite{Volpe:2006in}. It is therefore
essential to explore alternative avenues to get clues on this fundamental
question, such as indirect effects in dense environments like
core-collapse supernovae.

Core-collapse supernovae emit about 10$^{53}~$erg as neutrinos of all flavours during
their rapid gravitational collapse. Such neutrinos might play a role on the two major supernova unsolved problems,
namely understanding how the explosion finally occurs and where the nucleosynthesis of the heavy elements, produced during the {\it r}-process, takes place.
While neutrinos from a massive star were first observed during the SN1987A explosion,
future observations of (extra)galactic or relic supernova neutrinos will help unravelling
supernova physics and/or unknown neutrino properties. 
For example, in \cite{Schirato:2002tg,Tomas:2004gr,Fogli:2004ff,Kneller:2007kg} the imprint of the shock wave on the neutrino time signal is investigated; while avenues for extracting information on the third neutrino mixing angle are discussed in
\cite{Dighe:1999bi,Engel:2002hg}.
This searches require advances in the modelling of supernova
dynamics, of neutrino propagation in dense environments and of our knowledge on neutrinos.

Impressive developments are
currently ongoing in our understanding of neutrino propagation in dense matter.  
While solar experiments \cite{Davis:1964aa,Ahmad:2001an,Eguchi:2002dm,Arpesella:2007xf}  have beautifully confirmed the oscillation enhancement induced
by the coupling with matter - the Mikheev-Wolfenstein-Smirnov or MSW effect  \cite{Wolfenstein:1977ue,Mikheev:1986wj} -,
recent theoretical investigations have shown that 
the inclusion of the neutrino self-interactions in dense environments introducing a non-diagonal
neutrino refractive index \cite{Pantaleone:1992eq} gives rise to
a wealth of new phenomena, as first pointed out in \cite{Samuel:1993uw}.
Various regimes have been
identified : the synchronized one \cite{Pastor:2001iu,Duan:2005cp}, 
the bipolar oscillations \cite{Duan:2005cp,Hannestad:2006nj} and the spectral split phenomenon \cite{Raffelt:2007cb,Raffelt:2007xt}. Since numerical calculations become more involved, 
analytical treatments for the three flavour case are being proposed (see
e.g. \cite{Dasgupta:2007ws}). The importance of the loop corrections to the neutrino refractive
index, the $V_{\mu \tau}$ potential \cite{Botella:1986wy}, is underlined in \cite{EstebanPretel:2007yq}.
Moreover constraints on neutrino mixing from shock re-heating and the {\it r}-process nucleosynthesis, including the neutrino-neutrino interaction, are investigated in
\cite{Qian:1994wh,Sigl:1994hc}.
The impact of the neutrino-neutrino interaction on the electron fraction relevant for the {\it r}-process
is investigated in \cite{Pastor:2002we,Balantekin:2004ug}.

In a previous work \cite{Balantekin:2007es} we have investigated the CP effects on the neutrino fluxes, on the electron fraction in a supernova (relevant for the {\it r}-process) as well 
as the possible impact on the supernova neutrino signal in an observatory on Earth. In particular we have shown analytically that no effects can be found on the electron (anti)neutrino fluxes, when muon and tau neutrino have the same fluxes at the neutrinosphere; while significant effects are obtained numerically on the fluxes when they differ. 
The calculations in  \cite{Balantekin:2007es} are obtained considering
interaction with matter at tree level only.
In this work we explore for the first time CP violation effects on the neutrino propagation in dense environments
including the standard MSW effect, the neutrino self-interactions and the 
{\it V}$_{\mu \tau}$ refractive index. We first show analytically that if the muon and tau neutrinos interact similarly,
the electron (anti)neutrinos are not sensitive to the CP violating phase, even in presence of neutrino self-interactions.
This result is general and valid for any matter density profile and/or initial neutrino luminosity.
We present numerical results, obtained within the three flavour formalism, 
on the neutrino oscillation probabilities and fluxes within the star. In particular we show that
both the probabilities and the fluxes 
become sensitive to the CP violating phase if muon and tau neutrino 
interact differently with matter (e.g. because of loop corrections or physics beyond the Standard Model.)
The paper is structured as follows. Section II presents the theoretical framework for describing the
neutrino propagation including the coupling with matter as well as the neutrino-neutrino interaction.
Section III gives the analytical and numerical results. 
Conclusions are drawn in Section IV.

\section{Theoretical framework }
\noindent
In a dense environment the non-linear coupled neutrino evolution equations with neutrino self-interactions are given by ( we follow the formalism of Ref.\cite{Duan:2006an} ):
\begin{equation}
\label{e:hamiltonian}\label{e:1}
i{ d \over{dt}} \psi_{{\nu}_{\underline{\alpha}}} =[H_0 + H_m + H_{\nu \nu}] \psi_{{\nu}_{\underline{\alpha}}}
\end{equation}
where $ \psi_{{\nu}_{\underline{\alpha}}}$ denote a neutrino created at the neutrinosphere initially in a flavour state $\alpha =e, \mu,\tau$, $H_0 = U H_{vac} U^{\dagger}$ is the Hamiltonian describing the vacuum oscillations 
$H_{vac}=diag(E_1,E_2,E_3)$, $E_{i=1,2,3}$ being the energies of the neutrino mass eigenstates, and
$U$ the unitary Maki-Nakagawa-Sakata-Pontecorvo matrix
\begin{equation}
\label{e:2}
U = T_{23} T_{13} T_{12} = \left(\matrix{
     1 & 0 & 0  \cr
     0 &  c_{23}  & s_{23} \cr
     0 & - s_{23} &  c_{23} }\right)
 \left(\matrix{
     c_{13} & 0 &  s_{13} e^{-i\delta}\cr
     0 &  1 & 0 \cr
     - s_{13} e^{i\delta} & 0&  c_{13} }\right)
 \left(\matrix{
     c_{12} & s_{12} &0 \cr
     - s_{12} & c_{12} & 0 \cr
     0 & 0&  1 }\right) ,
\end{equation}
$c_{ij} = cos \theta_{ij}$ ($s_{ij} = sin \theta_{ij}$) with $\theta_{12},\theta_{23}$ and $\theta_{13}$ the three neutrino mixing angles. The presence of a Dirac $\delta $ phase in Eq.(\ref{e:2}) renders $U$ complex
and introduces a difference between matter and anti-matter.
The $U$ matrix relates the mass and the flavour basis
\begin{equation}
  \label{e:3}
  \psi_{{\nu}_{\underline{\alpha}}} = \sum_i U_{\alpha i} \psi_i .
\end{equation}
 The neutrino 
interaction with matter is taken into account through an effective Hamiltonian
which corresponds, at tree level, to the diagonal matrix $H_m=diag(V_c,0 ,0)$,
where the $V_c (x) = \sqrt{2} G_F  N_e (x)$ potential, due to the charged-current  interaction, depends
on the electron density $N_e (x)$ (note that the neutral current interaction introduces an overall phase only).
 
The neutrino self-interaction term is 
\begin{equation}
 \label{e:4}
H_{\nu \nu} = \sqrt{2} G_F
\sum_{\alpha} \sum_{\nu_{\alpha},\bar{\nu}_{\alpha}} \int
 \rho_{{\nu}_{\underline{\alpha}}} ({\bf q}')(1 - {\bf \hat{q}} \cdot {\bf \hat{q}'}) dn_{\alpha} dq' 
\end{equation}
where $G_F$ is the Fermi coupling constant, 
$\rho = \rho_{{\nu}_{\underline{\alpha}}}$  ($- \rho^*_{{\nu}_{\underline{\alpha}}}$) is 
the density matrix for neutrinos (antineutrinos) 
\begin{equation}\label{e:5}
 \rho_{{\nu}_{\underline{\alpha}}} = \left(\matrix{
     |\psi_{\nu_e}|^2 & \psi_{\nu_e} \psi_{\nu_{\mu}}^*&   \psi_{\nu_e} \psi_{\nu_{\tau}}^* \cr
     \psi_{\nu_e}^* \psi_{\nu_{\mu}}  &   |\psi_{\nu_{\mu}}|^2     &   \psi_{\nu_{\mu}} \psi_{\nu_{\tau}}^* \cr
      \psi_{\nu_e}^* \psi_{\nu_{\tau}} &   \psi_{\nu_{\mu}}^* \psi_{\nu_{\tau}} &   |\psi_{\nu_{\tau}}|^2   }\right) ,
\end{equation} 
${\bf q}$ (${\bf q'}$) denotes the
momentum of the neutrino of interest (background neutrino) and $dn_{\alpha}$ is
the differential number density.  In the single-angle approximation, that assumes that the neutrinos
are all emitted with the same angle, i.e.  $ \rho ({\bf q}) = \rho (q)$, Eq.(\ref{e:5}) reduces to
\begin{equation}\label{e:4bis}
H_{\nu \nu} ={ \sqrt{2} G_F \over {2 \pi R_{\nu}^2}} D(r/ R_{\nu}) \sum_{\alpha} \int [\rho_{{\nu}_{\underline{\alpha}}} (q') L_{{\nu}_{\underline{\alpha}}} (q') - \rho_{\bar{{\nu}}_{\underline{\alpha}}}^*(q')  L_{\bar{{\nu}}_{\underline{\alpha}}}(q')] dq'
\end{equation}
with the geometrical factor
\begin{equation}\label{e:4tris}
D(r/R_{\nu}) = {1 \over 2} [1 - \sqrt{1 - ({R_{\nu} \over r})^2 }]^2
\end{equation}
where the radius of the neutrino sphere is $R_{\nu} = 10 $km,
and
\begin{equation}\label{e:4quadris}
L_{{\nu}_{\underline{\alpha}}}(r,E_{\nu})= {L^0_{{\nu}_{\underline{\alpha}}} \over{T_{\nu_{\underline{\alpha}}}^3 \langle E_{\nu_{\underline{\alpha}}} \rangle F_2(\eta)}}{{E_{\nu_{\underline{\alpha}}}^2} \over{1 + \exp{(E_{\nu_{\underline{\alpha}}}/T_{\nu_{\underline{\alpha}}} - \eta)}}}
\end{equation}
where $F_2(\eta)$ is the Fermi integral, $L^0_{{\nu}_{\underline{\alpha}}}$ and $T_{\nu_{\underline{\alpha}}}$ are the luminosity and temperature at the neutrinosphere.

\section{CP effects in presence of the $\nu$-$\nu$ interaction and $V_{\mu \tau}$ refractive index}
\subsection{Analytical results}
\noindent
One way to study under which conditions the electron (anti)neutrino
survival probabilities depend upon the CP violating phase $\delta$, is to demonstrate that the $\delta$ dependence of the total Hamiltonian $H_T=H_0 + H_m + H_{\nu \nu}$ 
factorises as follows \cite{Balantekin:2007es}:
\begin{equation}\label{e:6}
\tilde{H}_T(\delta) = S \tilde{H}_T(\delta=0) S^{\dagger}
\end{equation}
in the $T_{23}$ basis which is:
\begin{eqnarray}
\label{e:12}
\tilde{\psi}_{\mu} &=& \cos{\theta_{23}} \psi_{\mu} -
\sin{\theta_{23}} \psi_{\tau}, \\
\tilde{\psi}_{\tau} &=& \sin{\theta_{23}} \psi_{\mu} +
\cos{\theta_{23}} \psi_{\tau}. \nonumber
\end{eqnarray}
Here the whole dependence is in the unitary diagonal matrix $S =
diag(1,1,e^{- i\delta})$.
In fact, it is straightforward to show that for any such Hamiltonian the
corresponding evolution operator also factorizes as
\begin{equation}\label{e:7}
\tilde{U} ({\delta}) = S^{\dagger} \tilde{U}(\delta=0) S  \leftrightarrow \tilde{H}(\delta) = S^{\dagger} \tilde{H}(\delta=0) S
\end{equation}
As a  consequence one can demonstrate that the electron (anti)neutrino survival
probabilities satisfy
$P (\nu_e \rightarrow \nu_e, \delta \neq 0)=P (\nu_e \rightarrow \nu_e, \delta = 0)$ and that the
appearance probabilities satisfy :
\begin{equation}
\label{e:8}
P (\nu_{\mu} \rightarrow \nu_e, \delta \neq 0) + 
P (\nu_{\tau} \rightarrow \nu_e, \delta \neq 0) =
P (\nu_{\mu} \rightarrow \nu_e, \delta =0) + 
P (\nu_{\tau} \rightarrow \nu_e, \delta =0) .
\end{equation}  
Since the $\nu_e$ ($\bar{\nu}_e$) fluxes are given by
\begin{equation}\label{e:9}
{\phi}_{\nu_{e}}(\delta) =  L_{\nu_{\underline{e}}}P(\nu_e \rightarrow \nu_e) + 
L_{\nu_{\underline{\mu}}}P(\nu_{\mu} \rightarrow \nu_e)+L_{\nu_{\underline{\tau}}}P(\nu_{{\tau}} \rightarrow \nu_e)
\end{equation}
where $L_{\nu_{\underline{x}}}$ are the neutrino fluxes at the neutrinosphere,
from Eqs.(\ref{e:8}-\ref{e:9}) one can see that 
$\phi_{\nu_e}$ ($\phi_{\bar{\nu}_e}$) are not sensitive to the CP violating phase 
if muon and tau neutrinos interact with matter in the same way (i.e. $ L_{\nu_{\underline{\mu}}}= L_{\nu_{\underline{\tau}}}$).
This is demonstrated in \cite{Balantekin:2007es} in the case of the standard MSW case only, i.e. $H =  H_0 + H_m$.
Note that a similar conclusion is drawn in \cite{Akhmedov:2002zj}, using a different procedure.
Besides, from Eqs (\ref{e:8},\ref{e:9}), one can see that if the condition $
L_{\nu_{\underline{\mu}}} \neq L_{\nu_{\underline{\tau}}}$ is relaxed then $\phi_{{\nu}_e}$ and $\phi_{\overline{\nu}_{e}}$ become dependent on delta, as first pointed out in \cite{Balantekin:2007es}.

\begin{figure}[t]
\vspace{.6cm}
\centerline{\includegraphics[scale=0.3,angle=0]{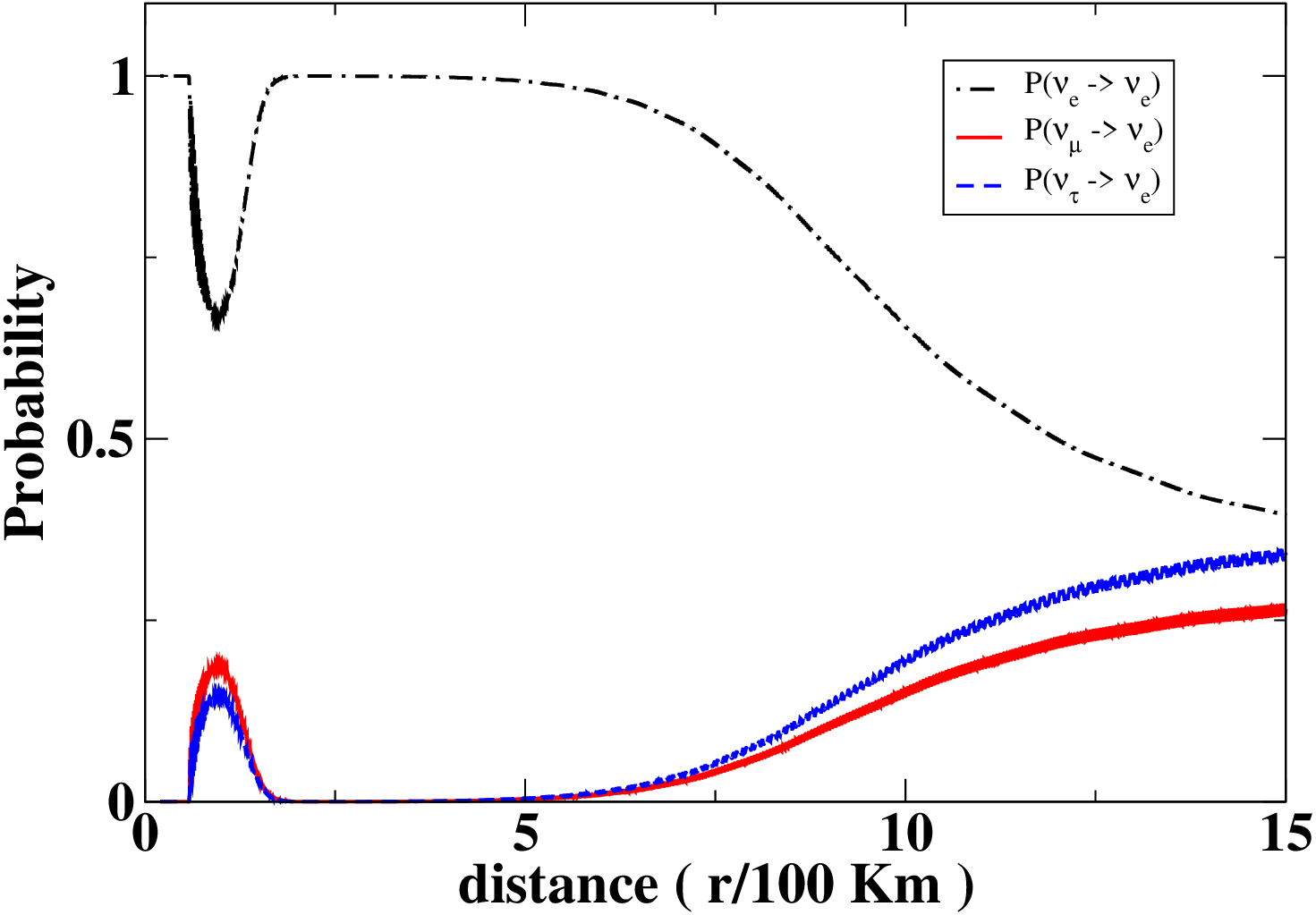}\hspace{.2cm}
\includegraphics[scale=0.3,angle=0]{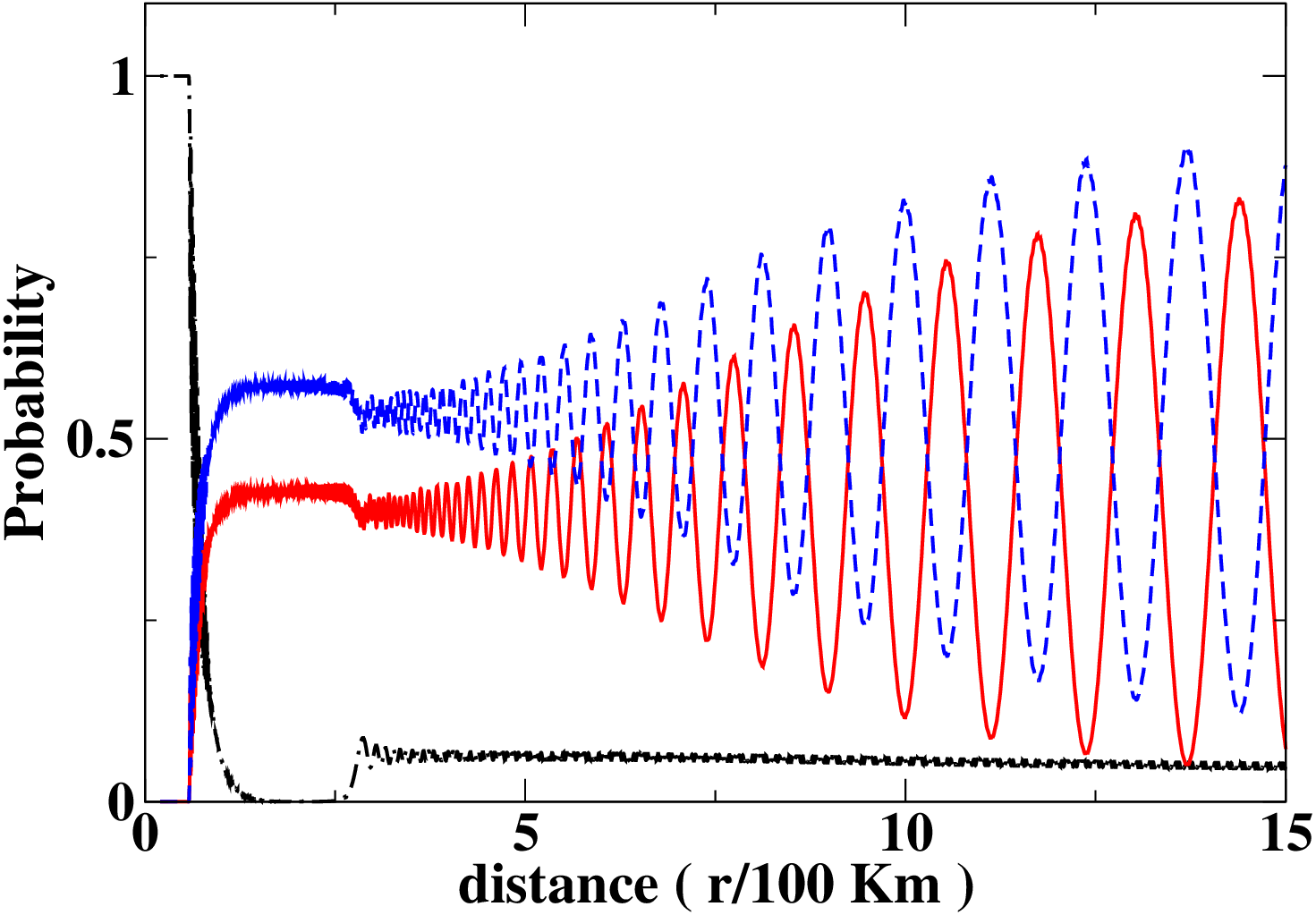}}
\caption{Neutrino (left) and antineutrino (right) oscillation probabilities in three flavours, as a function of the distance from the neutron-star surface (10 km), including the neutrino-neutrino interaction and $V_{\mu \tau}$ 
refractive index. The different 
curves correspond to electron (anti)neutrinos (dot-dashed), muon (solid) and tau (dashed) (anti)neutrinos.
The results are obtained solving Eqs.(\ref{e:1}-\ref{e:4tris}) numerically for a neutrino energy of 5 MeV as an example. The case of inverted hierarchy and small neutrino mixing angle $\theta_{13}$ is shown where the neutrino self-interaction effects are particularly impressive : the regimes of synchronized and bipolar oscillations can be recognised in the first 100 km. In the case of the electron neutrinos (left figure), the spectral split is also apparent. }
\label{fig:probnu3f}
\end{figure}

We now show that Eq.(\ref{e:6}) is indeed satisfied for the total Hamiltonian of Eq.(\ref{e:1}) including the non-linear $H_{\nu \nu}$ term of Eq.(\ref{e:4}).
Let us start with the Liouville-Von Neumann equation for the density matrix ($\hbar=1$) :
\begin{equation}\label{e:13a}
i{{d \rho_{{\nu}_{\underline{\alpha}}}(\delta)} \over{dt}} = [U H_{vac} U^{\dagger} + H_m + H_{\nu \nu}(\delta),\rho_{{\nu}_{\underline{\alpha}}}(\delta)] ,
\end{equation}
To prove our result (that the CP-violating phase can be factorized out of the total Hamiltonian which includes $H_{\nu \nu}$), one has to rotate in the $T_{23}$ basis, since the $S$ matrix contained in $T_{13}$ (which can be re written as $T_{13}=S^{\dagger}T_{13}^0S$ ) does not commute with $T_{23}$.
We then obtain :
\begin{equation}\label{e:13b}
i{{d S\tilde{\rho}_{{\nu}_{\underline{\alpha}}}(\delta)S^{\dagger}} \over{dt}} = [T_{13}^0T_{12}H_{vac}T_{12}^{\dagger}{T_{13}^0}^{\dagger}+H_m+S\tilde{H}_{\nu \nu}(\delta)S^{\dagger},S\tilde{\rho}_{{\nu}_{\underline{\alpha}}}(\delta)S^{\dagger}] , 
\end{equation}
where 
\begin{equation}\label{e:5a}
 \tilde{\rho}_{{\nu}_{\underline{\alpha}}} = \left(\matrix{
     P(\nu_{\alpha} \rightarrow \nu_e) & \psi_{\nu_e} \tilde{\psi}_{\nu_{\mu}}^*&   \psi_{\nu_e} \tilde{\psi}_{\nu_{\tau}}^* \cr
     \psi_{\nu_e}^* \tilde{\psi}_{\nu_{\mu}}  &   P(\nu_{\alpha} \rightarrow \tilde{\nu}_{\mu})     &   \tilde{\psi}_{\nu_{\mu}} \tilde{\psi}_{\nu_{\tau}}^* \cr
      \psi_{\nu_e}^* \tilde{\psi}_{\nu_{\tau}} &   \tilde{\psi}_{\nu_{\mu}}^* \tilde{\psi}_{\nu_{\tau}} &   P(\nu_{\alpha} \rightarrow \tilde{\nu}_{\tau})   }\right) .
\end{equation} 

Let us now consider the evolution equation of the linear combination
$ \sum_{\nu_{\alpha}}L_{\nu_{\underline{\alpha}}} S\tilde{\rho}_{{\nu}_{\underline{\alpha}}}({\bf q},\delta)S^{\dagger} $ at a given momentum ${\bf q}$.
At the initial time, this quantity reads, in the $T_{23}$ basis of Eq.(\ref{e:12}), as :
\begin{equation}\label{e:13c}
\sum_{\nu_{\alpha}}  L_{\nu_{\underline{\alpha}}}  S\tilde{\rho}_{{\nu}_{\underline{\alpha}}} ({\bf q},\delta, t=0)S^{\dagger} = \left(\matrix{
     L_{\nu_{\underline{e}}}  & 0 & 0  \cr
     0 &  c_{23}^2L_{\nu_{\underline{\mu}}}+s_{23}^2 L_{\nu_{\underline{\tau}}}  & c_{23}s_{23}e^{-i\delta}(L_{\nu_{\underline{\mu}}}-L_{\nu_{\underline{\tau}}}) \cr
     0 &  c_{23}s_{23}e^{i\delta}(L_{\nu_{\underline{\mu}}}-L_{\nu_{\underline{\tau}}}) &  s_{23}^2 L_{\nu_{\underline{\mu}}}+ c_{23}^2L_{\nu_{\underline{\tau}}}  }\right)
\end{equation} 

\begin{figure}
\vspace{.6cm}
\centerline{\includegraphics[scale=0.3,angle=0]{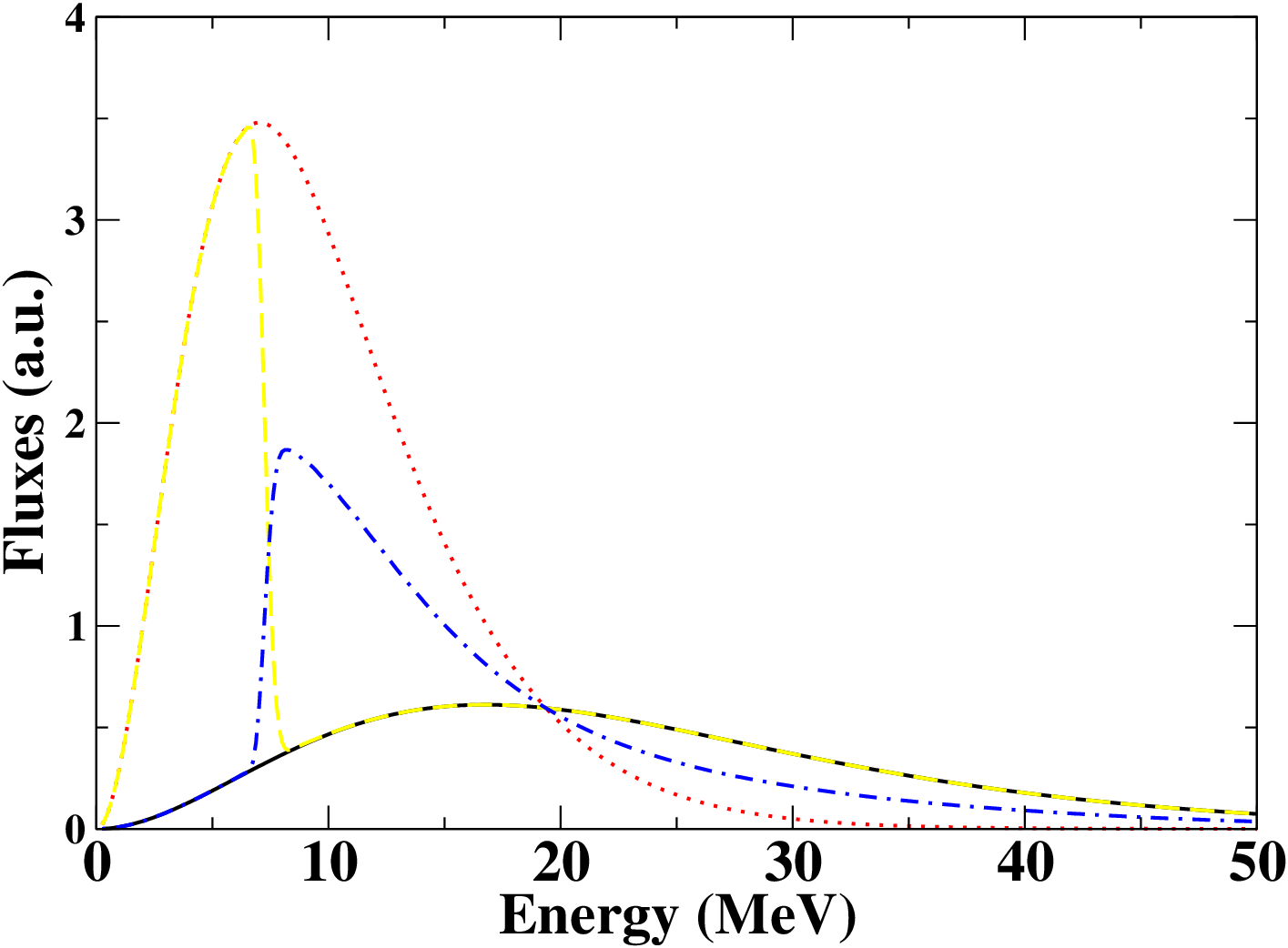}\hspace{.2cm}
\includegraphics[scale=0.3,angle=0]{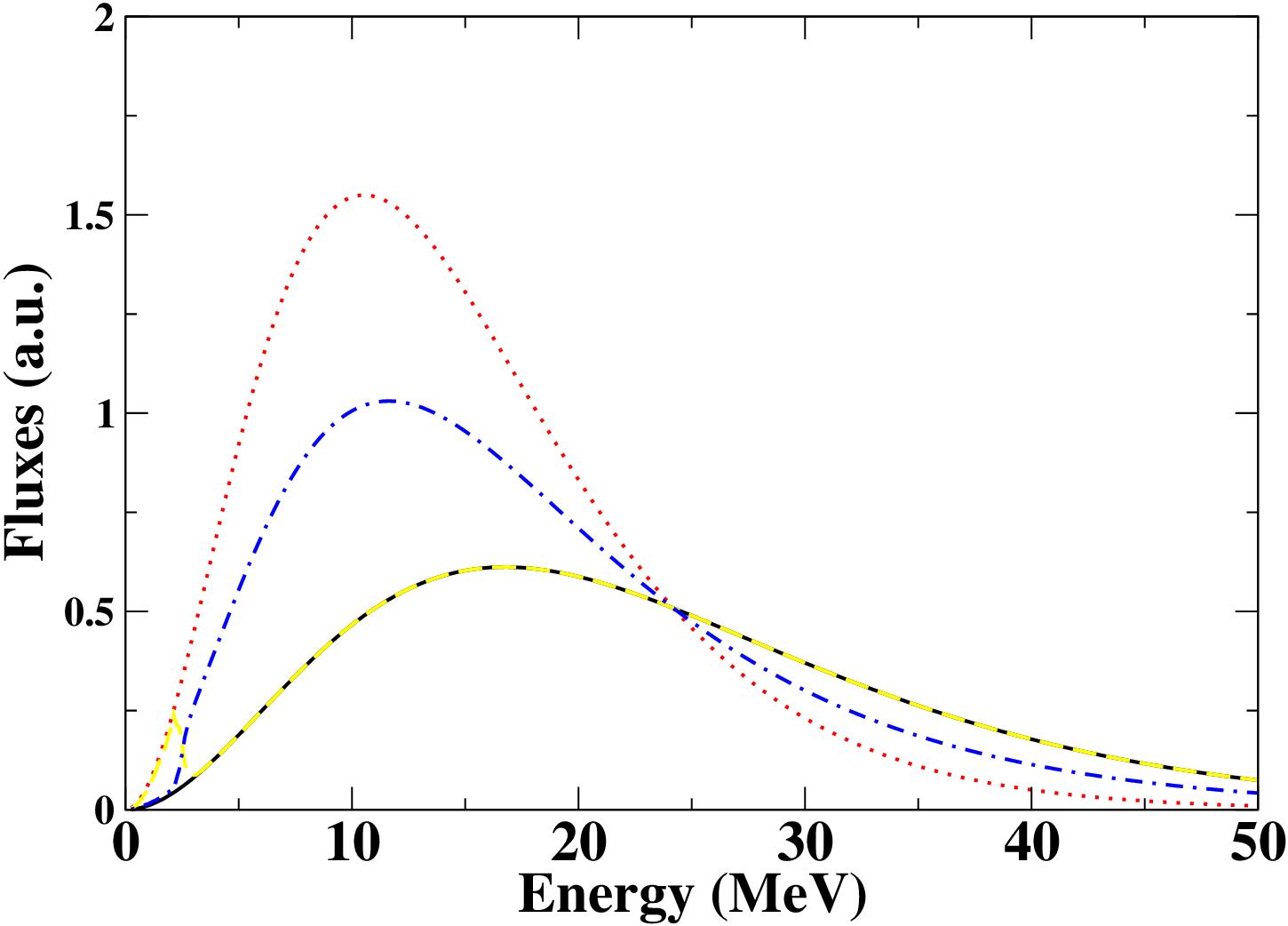}}
\caption{Neutrino (left) and anti-neutrino (right) spectra, at 200 km from the neutron-star surface. The different curves correspond to :
the original Fermi-Dirac distributions for $\nu_e$ (dotted) and $\nu_{\mu}$ (solid);  the $\nu_e$ (dashed) and  $\nu_{\mu}$ (dot-dashed) fluxes after the evolution in the star with the neutrino self-interaction.
The results are obtained for an inverted hierarchy and a small third neutrino mixing angle. While neutrinos show a spectral split, anti-neutrinos undergo full flavour conversion.}
\label{fig:flux3f}
\end{figure}
\noindent
One immediately sees that this quantity does not depend on $\delta$ if and only if $ L_{\nu_{\underline{\mu}}}= L_{\nu_{\underline{\tau}}}$.
Moreover, the total Hamiltonian of Eq.(\ref{e:13b}) is independent of $\delta$ at initial time since 
$T_{13}^0T_{12}H_{vac}T_{12}^{\dagger}{T_{13}^0}^{\dagger}+H_m$ does not depend on $\delta$ (at any time) and 

\begin{equation}\label{e:13c2}
S\tilde{H}_{\nu \nu}(t=0,\delta)S^{\dagger} = \sqrt{2} G_F
\sum_{\alpha} \int
  (1 - {\bf \hat{q}} \cdot {\bf \hat{q}'}) [S\tilde{\rho}_{{\nu}_{\underline{\alpha}}} (t=0,q')S^{\dagger} L_{{\nu}_{\underline{\alpha}}} (q') - S\tilde{\rho}_{\overline{{\nu}}_{\underline{\alpha}}}^*(t=0,q')S^{\dagger}  L_{\overline{{\nu}}_{\underline{\alpha}}}(q')] dq' 
\end{equation}
is equal to $\tilde{H}_{\nu \nu}(t=0,\delta=0)$ initially when $ L_{\nu_{\underline{\mu}}}= L_{\nu_{\underline{\tau}}}$(and $ L_{\overline{\nu}_{\underline{\mu}}}= L_{\overline{\nu}_{\underline{\tau}}}$).
In that case, one can see by recurrence from the Liouville-Von Neumann equation Eq.(\ref{e:13a}) that the evolution of the term $ \sum_{\nu_{\alpha}}L_{\nu_{\underline{\alpha}}} S\tilde{\rho}_{{\nu}_{\underline{\alpha}}} ({\bf q}, \delta)S^{\dagger} $ is exactly the same as the term $ \sum_{\nu_{\alpha}}L_{\nu_{\underline{\alpha}}} \tilde{\rho}_{{\nu}_{\underline{\alpha}}} ({\bf q},\delta=0) $, since they have the same initial conditions (for any ${\bf q}$) and the same evolution equations.
Indeed, the exact same relation applying at the same time for the anti-neutrino case 
(where the sign of $\delta$ has to be changed), one simultaneously obtains that at any time:

\begin{equation}\label{e:13d}
\tilde{H}_{\nu \nu}(\delta) = S \tilde{H}_{\nu \nu}(\delta=0) S^{\dagger},
\end{equation}
hence :

\begin{equation}\label{e:13e}
\tilde{H}_T(\delta) = S \tilde{H}_T(\delta=0) S^{\dagger}.
\end{equation}
Note that the derivation holds both for the multi-angle case Eq.(\ref{e:4}) and the single-angle case Eq.(\ref{e:4bis}).

This implies that the (anti-)electron neutrino survival probability is independent of $\delta$ and Eq.(\ref{e:8}) is valid, therefore 
$\phi_{\nu_e}(\delta)=\phi_{\nu_e}(\delta=0)$ and
$\phi_{\overline{\nu}_e}(\delta)=\phi_{\overline{\nu}_e}(\delta=0)$, even
considering the presence of neutrino-neutrino interaction, if muon and tau
neutrino fluxes at the neutrinosphere are equal.
When the fluxes $ L_{\nu_{\underline{\mu}}} $ and $ L_{\nu_{\underline{\tau}}}$ are different, the derivation does not hold anymore, since 
$ \sum_{\nu_{\alpha}}L_{\nu_{\underline{\alpha}}} S\tilde{\rho}_ {{\nu}_{\underline{\alpha}}} ({\bf q}, \delta)S^{\dagger} $ initially depend on $\delta$.

In the case where radiative corrections to the neutrino scattering are considered, there is an extra
term in $H_m = diag(V_c,0,V_{\mu \tau})$. For the corresponding Hamiltonian the $\delta$ dependence
cannot be factorized anymore and in general nothing prevents the electron (anti)neutrino survival
probabilities to be sensitive to the CP violating phase\footnote{Note that the possible inclusion of nonstandard neutrino interactions in the flavour neutrino mixing as e.g. in \cite{EstebanPretel:2007yu} implies that Eq.(9) does not hold anymore.}. 
These cases will be studied in the following.

\begin{figure}[t]
\vspace{.6cm}
\centerline{\includegraphics[scale=0.3,angle=0]{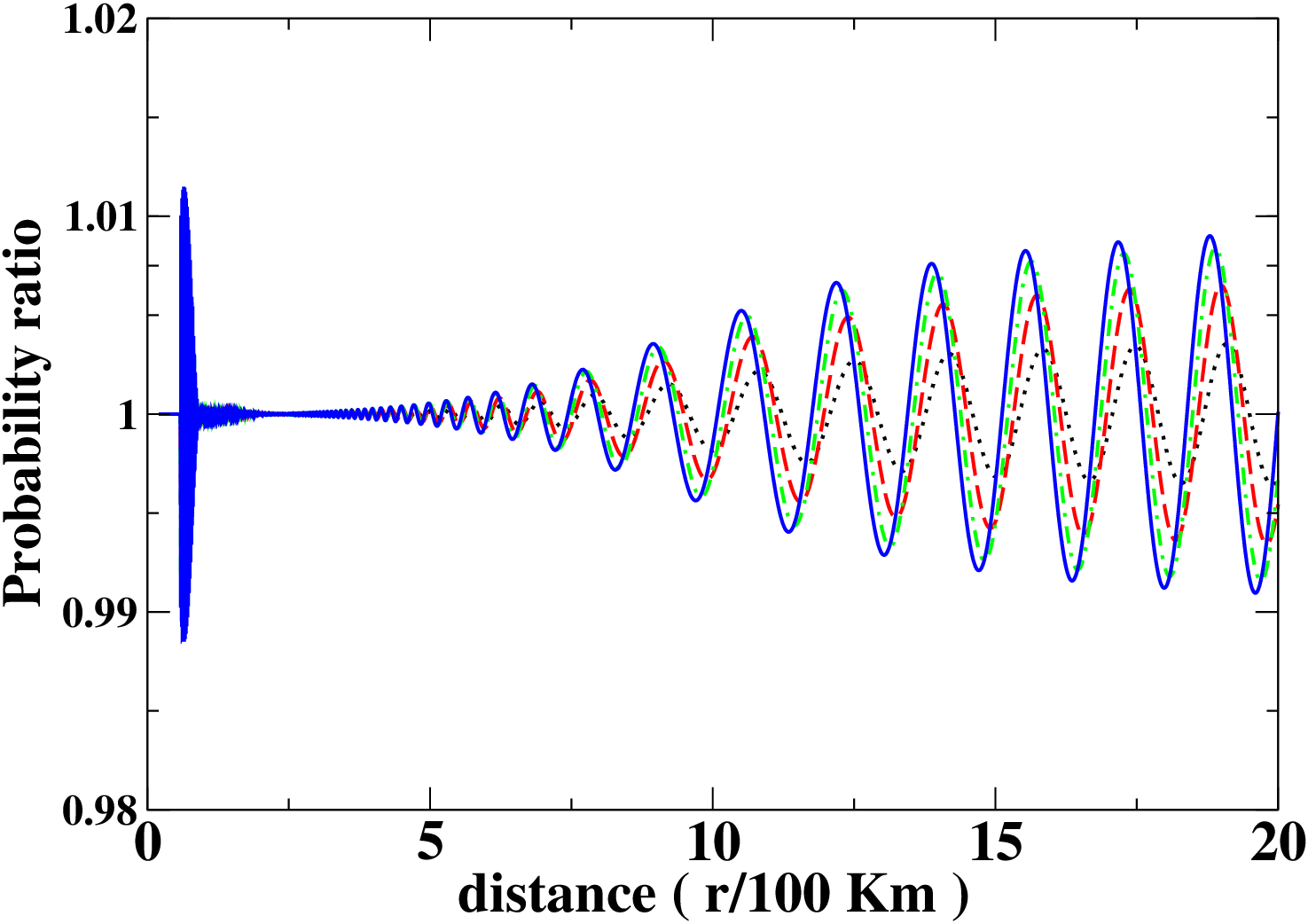}\hspace{.2cm}
\includegraphics[scale=0.3,angle=0]{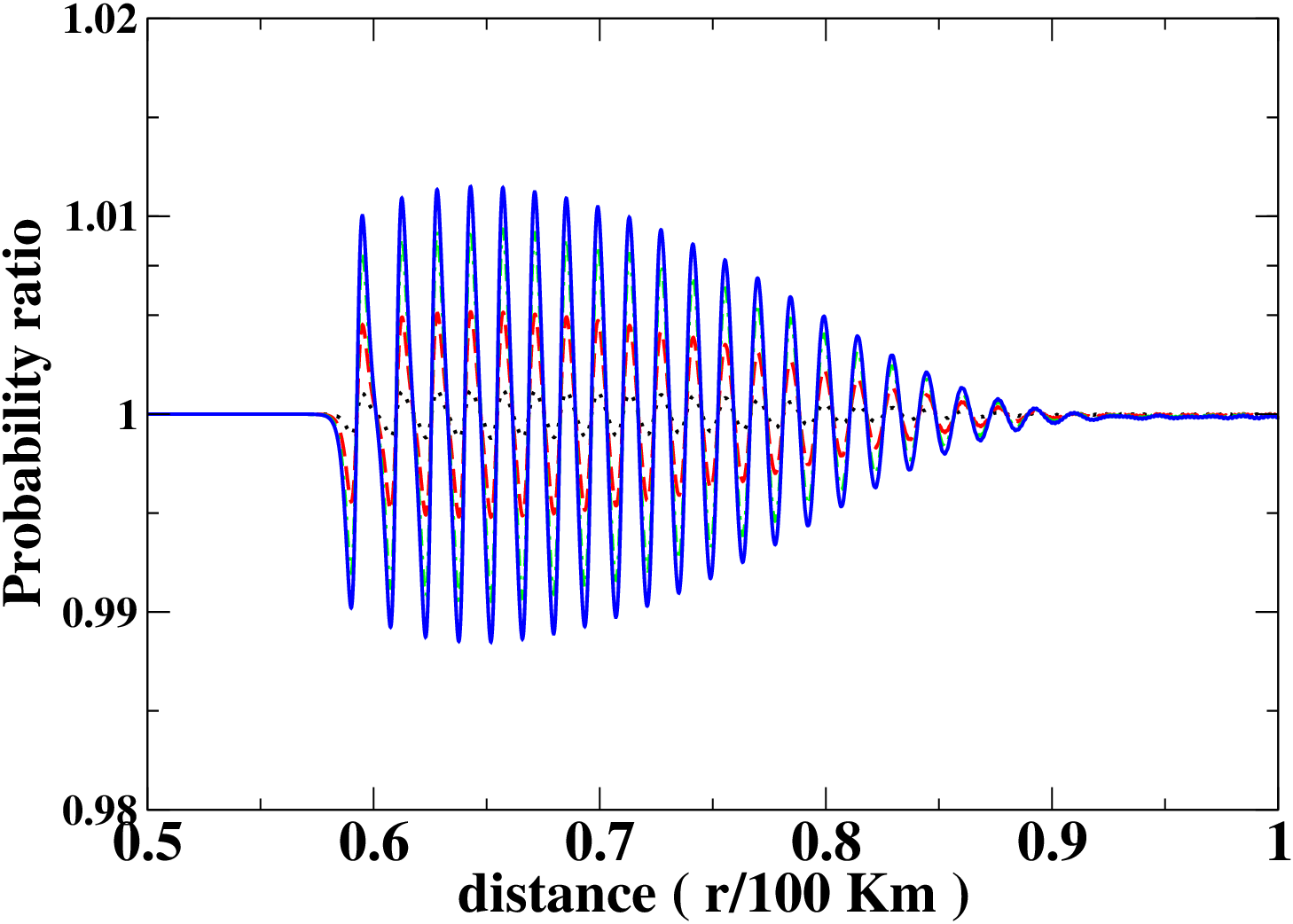}}
\caption{Ratios of the electron neutrino oscillation probabilities for a CP violating phase 
$\delta=45^{\circ}$ (dotted), $90^{\circ}$ (dashed), $135^{\circ}$ (dot-dashed), 180$^{\circ}$ (solid) over $\delta= 0^{\circ}$, as a function of the distance from the neutron star surface. The left figures shows the ratios up to 2000 km while the right figure 
presents the region between 50 to 100 km where collective effects induced by the neutrino self-interaction are maximal.
The results correspond to the case
of inverted hierarchy and small third neutrino mixing angle, for a neutrino energy of 5 MeV. }
\label{fig:probcas}
\end{figure}

\subsection{Numerical results}
\noindent
The main goal of this section is to investigate numerically the effects that can arise when the factorization Eq.(\ref{e:6}) is not satisfied. In particular this occurs in three cases : (i) if the MSW $H_0 + H_m$ Hamiltonian 
does not satisfy Eq.(\ref{e:7}) because the $V_{\mu \tau} $ refractive index is included; (ii) 
if the initial conditions Eq.(\ref{e:13c}) of the $H_{\nu \nu}$ Hamiltonian
are not independent on $\delta$, that happens when $ L_{\nu_{\underline{\mu}}} \neq L_{\nu_{\underline{\tau}}}$, which implies that the neutrino self-interaction term does not follow  Eq.(\ref{e:13d});
(iii) when (i) and (ii) occur. Therefore the results that we present here correspond to the following possibilities :
\begin{itemize}
\item [{\it a)}] $H_{\nu \nu} \neq 0$ and $V_{\mu \tau} \neq 0$ with the condition $ L_{\nu_{\underline{\mu}}}= L_{\nu_{\underline{\tau}}}$;
\item [{\it b)}]  $H_{\nu \nu} \neq 0$ and $V_{\mu \tau} =  0$ with the condition $ L_{\nu_{\underline{\mu}}} \neq L_{\nu_{\underline{\tau}}}$;
\item [{\it c)}] $H_{\nu \nu} \neq 0$ and $V_{\mu \tau} \neq 0$ with the condition  $ L_{\nu_{\underline{\mu}}} \neq L_{\nu_{\underline{\tau}}}$ .
\end{itemize}
It is important to note that a new feature arises in the {\it a), b)} and {\it c)} cases  : 
the  $P (\nu_e \rightarrow \nu_e)$ becomes dependent on $\delta$.
(It is a well known fact that the electron survival probability  does not depend on CP violation 
in vacuum and in presence of the interaction of matter at tree level.) 
In the following we investigate the CP effects by varying $\delta \in [0^{\circ}, 180^{\circ}]$ while we show
results for $\delta=180^{\circ}$ when at such value the effects are maximal.

The numerical results we present are obtained by solving the three flavour evolution equation of
Eq.(\ref{e:1})
 with a supernova density profile having a $1/r^{3}$ behavior  
that fits the numerical simulations shown in \cite{Balantekin:2004ug}.
Note that with such density profile the region of the first 100 km, where the neutrino self-interaction dominates, is well separated from the one of the MSW (high and low) resonances, produced by the interaction with ordinary matter. 

The oscillation
parameters are fixed at the present best fit values \cite{Amsler:2008zz}, 
namely
$\Delta m^2_{12}= 8 \times 10^{-5}$eV$^2$, sin$^2 2\theta_{12}=0.83$ and
$\Delta m^2_{23}= 3 \times 10^{-3}$eV$^2$, sin$^2 2\theta_{23}=1$ for
the solar and atmospheric differences of the mass squares and
mixings, respectively. For the third still unknown neutrino mixing angle
$\theta_{13}$, we take either the present upper limit 
sin$^2 2\theta_{13}=0.19$ at 90 $\%$ C.L. ({\it L}) or  a very small value of 
sin$^2 2\theta_{13}=3 \times 10^{-4}$ ({\it S})
 that might be attained at the future (third generation)  long-baseline
experiments \cite{Volpe:2006in}.  
To include the neutrino-neutrino interaction we use the single-angle
approximation of Eqs.(\ref{e:4bis}-\ref{e:4tris}) with $R_{\nu} = 10$ km, 
considering that all neutrinos are emitted radially.
The neutrino fluxes at the neutrinosphere $L_{\nu_{\underline{\alpha}}}$  
Eq.(\ref{e:4quadris}) are taken as Fermi-Dirac distributions with typical
average energies of $ <E_{{\nu}_e} >$ = 10 MeV, $ <E_{\bar{\nu}_e}>$ = 15 MeV and
$<E_{\nu_x}>$= 24 MeV with $\nu_x=\nu_{\mu},\nu_{\tau},\bar{\nu}_{\mu},\bar{\nu}_{\tau}$, unless
stated otherwise (the chemical potentials are assumed to be zero for simplicity). We take the neutrino luminosity
$L^0_{{\nu}_{\underline{\alpha}}}= 10^{51}~$erg $\cdot$ s$^{-1}$.

\begin{figure}[t]
\vspace{.6cm}
\centerline{\includegraphics[scale=0.3,angle=0]{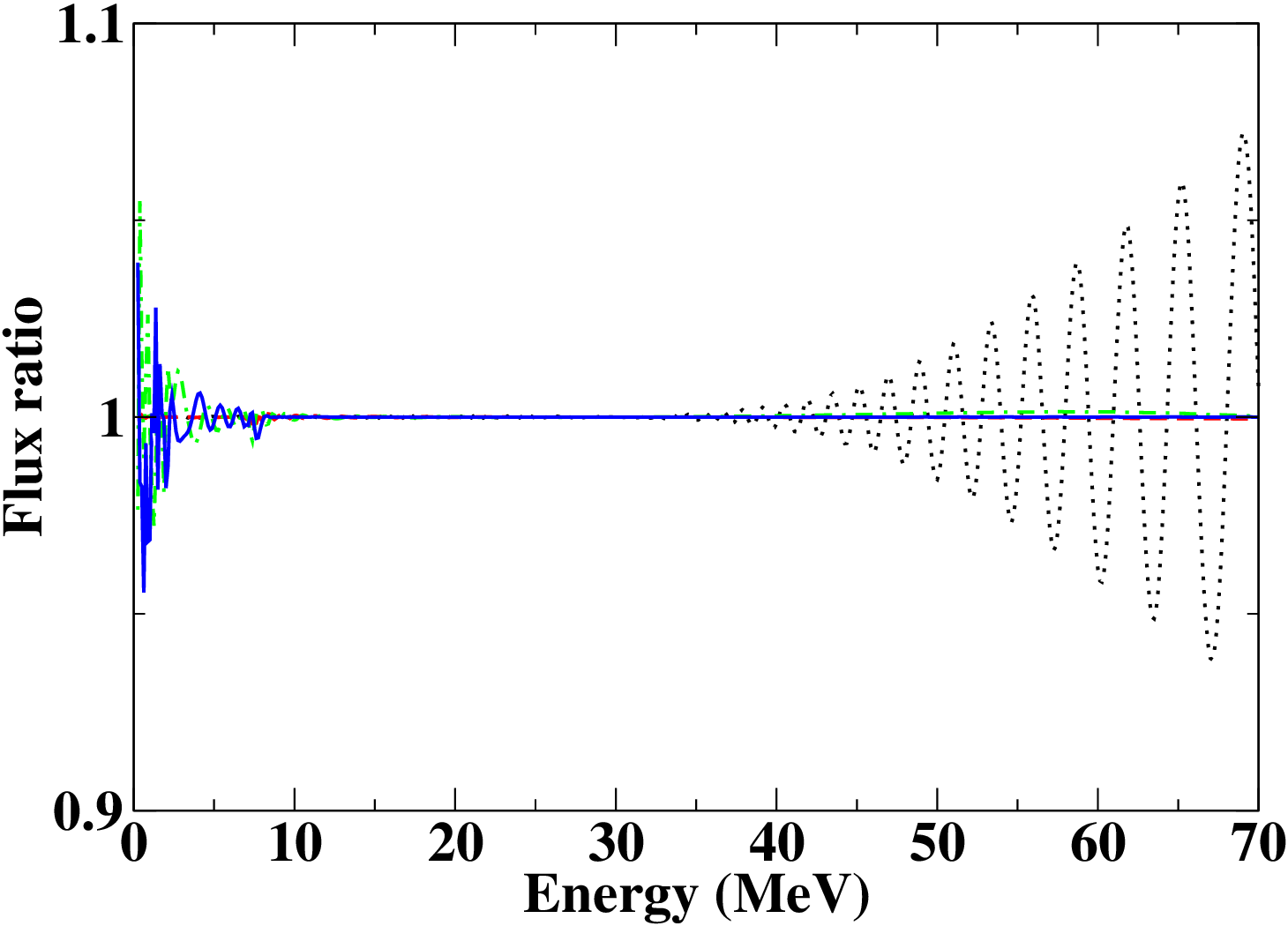}}
\caption{Ratios of the $\nu_e$ fluxes for a CP violating phase $\delta=180^{\circ}$ over $\delta= 0^{\circ}$
as a function of neutrino energy, at 1000 km within the star. The curves correspond to 
the following cases :
$H_{\nu \nu} = 0$ and $V_{\mu \tau} = 0$ (dotted), 
$H_{\nu \nu} \neq 0$ and $V_{\mu \tau} = 0$ (dashed) and
$H_{\nu \nu} \neq 0$ and $V_{\mu \tau} \neq 0$ (solid).
These are obtained with $L_{\nu_{\underline{\mu}}} \neq L_{\nu_{\underline{\tau}}}$ e.g. $T_{\nu_{\mu}} = 1.05~T_{\nu_{\tau}}$. The case    
$H_{\nu \nu} \neq 0$ and $V_{\mu \tau} \neq 0$ (dot-dashed)
with $L_{\nu_{\underline{\mu}}} = L_{\nu_{\underline{\tau}}}$ is also shown.
The results correspond to an inverted hierarchy and a small $\theta_{13}$. }
\label{fig:difcas}
\end{figure}

Our numerical results in three flavours present the collective oscillations induced by the neutrino-neutrino interaction,
already discussed in the literature (see e.g. \cite{Samuel:1993uw,Sigl:1992fn,Pastor:2001iu,Duan:2005cp,Hannestad:2006nj,Raffelt:2007cb,Raffelt:2007xt,EstebanPretel:2007yq}).  
Figure \ref{fig:probnu3f} presents the (anti)neutrino oscillation probabilities within the star. One recognizes the
synchronized regime in the first 50 km outside the neutrinosphere $R_{\nu}$
(assumed here to be equal to the neutron-star surface).
In this regime the strong neutrino-neutrino interaction makes neutrinos of all energies oscillate with the same frequency so that
flavour conversion is frozen, as
 discussed e.g. in \cite{Pastor:2001iu,Duan:2006jv}.
When the neutrino self-interaction term becomes smaller, the ordinary matter term starts to dominate producing large
bipolar oscillations (between 50 and 80 km) that produce strong flavour conversion for both neutrinos and anti-neutrinos, in particular for the case of inverted hierarchy, independently of the $\theta_{13}$ value \cite{Hannestad:2006nj}.
Finally neutrinos show complete (no)  flavour 
conversion for energies larger (smaller)  than a characteristic energy $E_c = 7.4$~MeV, due to lepton number conservation \cite{Raffelt:2007cb}. This is known as the spectral split phenomenon (apparent  around 150 km on Figure \ref{fig:probnu3f}, left). (Note that the effect of the partially non-adiabatic MSW low resonance can be seen
at around 270 km of Figure \ref{fig:probnu3f}, right.)

The neutrino-neutrino interaction might have an important impact on the neutrino spectra as well.
If in the case of normal hierarchy the flavor evolution of both electron neutrinos and anti-neutrinos are essentially the same as in the case where matter only is included, for the case of inverted hierarchy, important modifications are found compared to the MSW case \cite{Fogli:2007bk}. 
While electron neutrinos swap their spectra with muon and tau neutrinos (Figure \ref{fig:flux3f});
the electron anti-neutrinos show a complete spectral swapping (Figure \ref{fig:flux3f}).
Such behaviours are found for both large and small values of the third neutrino mixing angle, in constrast with 
the standard MSW effect. (Note that anti-neutrinos of energies less than 2 MeV have already undergone the MSW low resonance at 200 km, as can be seen from Figure \ref{fig:flux3f}, right.)

Let us now discuss the CP violation effects in presence of the neutrino-neutrino interaction\footnote{Note that a comment is made in \cite{Duan:2007sh,Duan:2008za} on the $\delta$ effects on the neutrino fluxes in the presence of the neutrino self-interaction in a core-collapse supernova.} and of the loop corrections to the neutrino refractive index, with the condition that the muon and tau fluxes at the neutrinosphere are equal ($L_{\nu_{\mu}}=L_{\nu_{\tau}}$). 
Figure \ref{fig:probcas} shows the ratios of the electron neutrino oscillation probabilities for different $\delta $ values, as a function of the distance within the star. A 5 MeV neutrino is taken, as an example. One can see that the $\delta$ effects are at the level of 1 $\%$. Note that the presence of $H_{\nu \nu}$ with $V_{\mu \tau}$ amplifies these effects that are at the level of less than $0.1 \%$ and smaller, when $V_{\mu \tau}$ only is included\footnote{Note that it was first pointed out in \cite{Yokomakura:2002av} that the inclusion of the $V_{\mu \tau}$ refractive index renders the electron neutrino survival probability slightly $\delta$ dependent.  }. One can also see that in the synchronized regime the CP effects are "frozen" while they develop with the bipolar oscillations. Similar modifications are also found in the case of electron anti-neutrinos, with effects up to $10 \%$ for low energies (less than 10 MeV). Note that the latter might be partially modified in a multi-angle calculation, since it has been shown that the decoherence effects introduced by multi-angles modify the electron anti-neutrino energy spectra in particular at low energies \cite{Fogli:2007bk}. Multi-angle decoherence is also discussed in 
\cite{Duan:2006an,Duan:2006jv,EstebanPretel:2007ec,EstebanPretel:2008ni}. 
To predict how these effects modify the numerical results presented in this paper 
would require a full multi-angle calculation.

\begin{figure}[t]
\vspace{.6cm}
\centerline{\includegraphics[scale=0.3,angle=0]{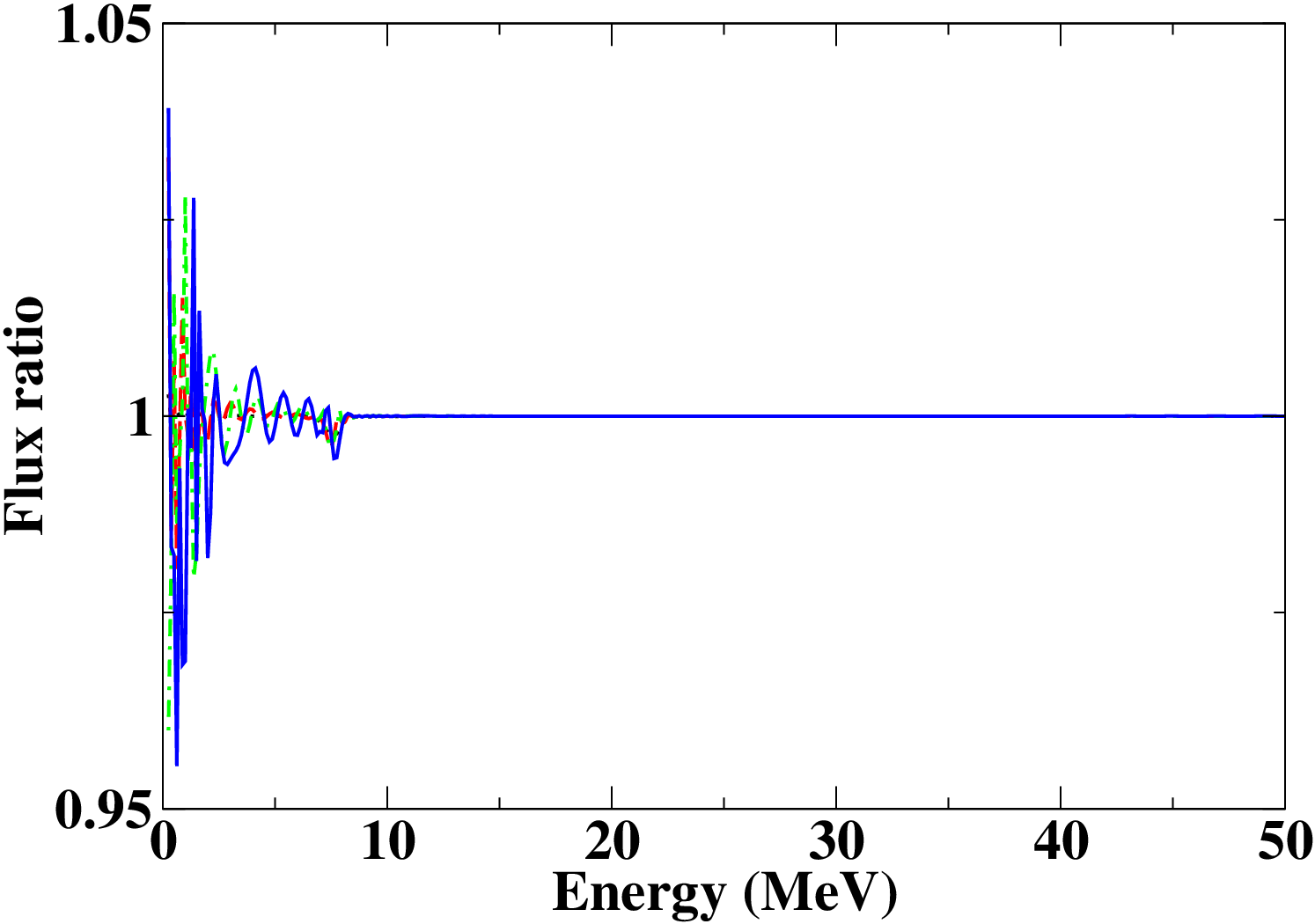}\hspace{.2cm}
\includegraphics[scale=0.3,angle=0]{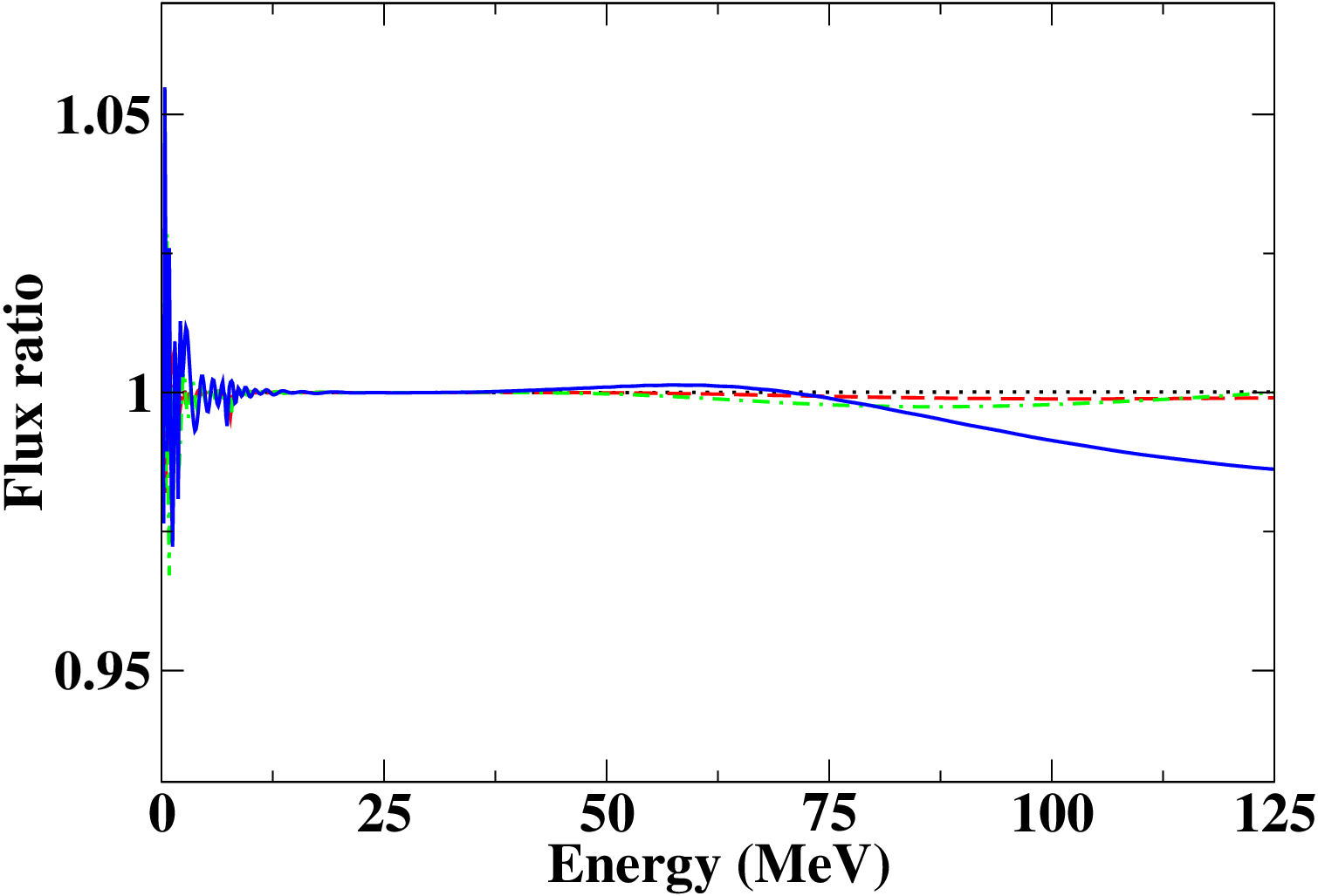}}
\caption{Ratios of the $\nu_e$ fluxes for a CP violating phase $\delta=180^{\circ}$ over $\delta= 0^{\circ}$
as a function of neutrino energy. They correspond to 
inverted hierarchy and small $\theta_{13}$ and 
different distances from the neutron star surface, 
i.e. 200 km (dotted), 500 km (dashed), 750 (dot-dashed), 1000 (solid). The results 
include the $\nu$-$\nu$ interaction and the $V_{\mu \tau}$ refractive index. They are obtained using
equal $\nu_{\mu}$ and $\nu_{\tau}$ fluxes at the neutrinosphere (left) or taking  
$T_{\nu_{\mu}}=1.05~T_{\nu_{\tau}}$ (right). For the $\bar{\nu}_e$ fluxes 
deviations up to 10$\%$ are found at energies lower than 10 MeV.}
\label{fig:nueIHSA}
\end{figure}

The modifications induced by $\delta$ on the electron neutrino fluxes are shown in Figure \ref{fig:difcas}
for the {\it a)}, {\it b)}, and {\it c)} cases, in comparison with a calculation within the MSW effect at tree level only as investigated in a previous work \cite{Balantekin:2007es}.
Figure \ref{fig:nueIHSA} shows how the CP effects evolve as a function of the distance from the neutron-star surface for the {\it a)} and {\it c)} cases. 
To differentiate the muon and tau neutrino fluxes at the neutrinosphere here we take as an example
$T_{\nu_{\mu}}=1.05~T_{\nu_{\tau}}$ (note that in \cite{Balantekin:2007es} differences of $10 \%$ are considered). In general, we have found that the inclusion of the neutrino self-interaction
in the propagation reduces possible effects from $\delta$ compared to the case without neutrino-neutrino interaction, as can be seen in Figure \ref{fig:difcas}.
In all studied cases both for $\nu_e$ and $\bar{\nu}_e$ we find effects up to a few percent
at low neutrino energies, and at the level of $0.1 \%$ at high energies (60 - 120 MeV).
Our numerical results show deviation at low energies that can sometimes be larger than in absence of neutrino self-interaction (Figure \ref{fig:difcas}), those at high energies turn out to be much smaller. This effect of the neutrino-neutrino interaction might be due to the presence of the synchronized regime that freezes possible flavour conversion at initial times and therefore also reduces the modifications coming from a non zero CP violating phase at later times.
 
\section{Conclusions}
We have investigated possible effects of the CP violating phase on the neutrino propagation in
dense matter when interaction with matter without/with loop corrections and the neutrino self-interaction 
are included. Our analytical results demonstrate that, at tree level, 
even when the neutrino-neutrino interaction is included 
there are no CP violating effects on the electron (anti)neutrino fluxes
in the star unless $\nu_{\mu}$ and $\nu_{\tau}$ fluxes differ at the neutrinosphere. 
If such condition is not satisfied, a totally new feature arise, namely that the electron 
(anti)neutrino oscillation probabilities (and fluxes) become sensitive to the CP violating phase $\delta$.
The latter is also true when the loop corrections to the refractive index are included.
We find numerically that, in most cases studied, the modifications introduced by the CP violating phase are larger (smaller) at low (high) energies than in the case where the neutrino-neutrino interaction is not included, and at the level of a few percent. We also find numerically that, even assuming that the muon and tau neutrinos have the same fluxes at the neutrinosphere, the CP effects induced by $V_{\mu\tau}$ only are amplified by the neutrino self-interactions up to several percent.

\vspace*{0.5cm}
The authors acknowledge the support from "Non standard
neutrino  properties and their impact in astrophysics and cosmology", Project No. ANR-05-JCJC-0023.

\textit{}

\end{document}